\documentclass[12pt,twoside]{article}
\usepackage{fleqn,espcrc1}
\usepackage{epsfig}
\title{Charged particle flow measurement for $|\eta|<5.3$ with the PHOBOS detector} 

\author{Inkyu Park for the PHOBOS Collaboration\\[0.2cm]
\small
B.B.Back$^1$, M.D.Baker$^2$, 
D.S.Barton$^2$, R.R.Betts$^{1,6}$, R.Bindel$^7$,  
A.Budzanowski$^3$, W.Busza$^{4}$, A.Carroll$^2$,
M.P.Decowski$^4$, 
E.Garcia$^7$, N.George$^1$, K.Gulbrandsen$^4$, 
S.Gushue$^2$, C.Halliwell$^6$, J.Hamblen$^8$, 
G.A.Heintzelman$^2$, C.Henderson$^4$, R.Ho\l y\'{n}ski$^3$, D.J.Hofman$^6$,
B.Holzman$^6$, 
E.Johnson$^8$, J.L.Kane$^4$, J.Katzy$^4$, N. Khan$^8$, W.Kucewicz$^{6}$,
P.Kulinich$^4$,
W.T.Lin$^5$, S.Manly$^8$,  D.McLeod$^6$, J.Micha\l owski$^3$,
A.C.Mignerey$^7$, J.M\"ulmenst\"adt$^{4}$,R.Nouicer$^6$, 
A.Olszewski$^{2,3}$, R.Pak$^2$, I.C.Park$^8$, 
H.Pernegger$^4$, C.Reed$^4$, L.P.Remsberg$^2$, 
M.Reuter$^6$, C.Roland$^4$, G.Roland$^4$, L.Rosenberg$^4$, 
P.Sarin$^4$, P.Sawicki$^3$, 
W.Skulski$^8$, 
S.G.Steadman$^4$, 
G.S.F.Stephans$^4$, P.Steinberg$^2$, M.Stodulski$^3$, A.Sukhanov$^2$, 
J.-L.Tang$^5$, R.Teng$^8$, A.Trzupek$^3$, 
C.Vale$^4$, G.J.van Nieuwenhuizen$^4$, 
R.Verdier$^4$, B.Wadsworth$^{4}$, F.L.H.Wolfs$^8$, B.Wosiek$^3$, 
K.Wo\'{z}niak$^3$, 
A.H.Wuosmaa$^1$, B.Wys\l ouch$^4$ 
\\[0.2cm]
$^1$~Argonne National Laboratory, 
$^2$~Brookhaven National Laboratory, 
$^3$~Institute of Nuclear Physics, Krak\'{o}w, Poland
$^4$~Massachusetts Institute of Technology, 
$^5$~National Central University, Chung-Li, Taiwan
$^6$~University of Illinois at Chicago, 
$^7$~University of Maryland, 
$^8$~University of Rochester 
}

\begin{document}

\maketitle

\begin{abstract}\small
The measurement of elliptic flow for charged particles in Au-Au collisions at $\sqrt{s_{_{NN}}} = 130$ GeV using the PHOBOS detector at the Relativistic Heavy Ion Collider (RHIC) is presented.  
Charge particle distributions over a wide pseudo-rapidity region ($|\eta|<5.3$) are measured by the PHOBOS silicon multiplicity detectors.
These distributions are used to extract the final-state azimuthal anisotropy.
The centrality and pseudo-rapidity dependences of elliptic flow, averaged over momenta on an event-by-event basis and particle species, are discussed.
\end{abstract}

\section{Introduction}

Collective flow in ultra-relativistic heavy ion collisions has become an important tool for searching for a hadronic phase transition since it is sensitive to the early stages of the collision system where the phase transition might occur, and its magnitude provides information on the Equation of State and the degree of thermalization of the system.
In addition, the reaction plane information, a by-product of the flow measurement, will allow a better understanding in other physics measurements, such as HBT correlation or particle spectra.

Flow is quantified as the coefficients of the Fourier expansion of the azimuthal angle distribution of particles with respect to the reaction plane: ${dN / d(\phi-\Psi_R)}= N_0(1+\sum_n 2v_n \cos[n(\phi-\Psi_R)])$, where $\Psi_R$ denotes the reaction plane angle, and the first two coefficients, $v_1$ and $v_2$, are the strengths of the directed flow and the elliptic flow, respectively.
Both types of flow have been observed and extensively studied in lower energy experiments~\cite{ref:c.ogilvie&m.lisa}.

\section{Detector and data set}

Details of the PHOBOS detector can be found in Ref.~\cite{ref:r.pak}.
The parts of the detector relevant for this analysis are, two sets of sixteen scintillator paddle counters (Paddle) for the event trigger and centrality measurement, two silicon tracking devices (Spectrometers) used for the event vertex reconstruction, the octagonal silicon pad detector (Octagon) and the annular silicon pad detectors (Rings) used to determine charged particle multiplicity. 

Collision events were selected by a trigger based on the coincidence of two or more hits in both Paddle counters. 
The signal in the Paddle counters increases with centrality and therefore can be used as a measure of event centrality.
Further details on the trigger and on the centrality determination can be found in Ref.~\cite{ref:j.katzy}.

Reconstruction of the event vertex, {\it i.e.} the beam-beam collision point, starts with straight-line tracking based on a simple $\chi^2$ fitting method in the first six planes of both Spectrometers where the magnetic field is negligible.
The event vertex is then determined as the common vertex of the straight line tracks.

Once the event vertex is known, hit distributions in the Octagon and Ring detectors are translated into particle pseudo-rapidity and azimuthal angle distributions.
For collisions near the nominal interaction point, the Octagon and Rings cover $|\eta|<5.3$.
Hits are defined as pads having more than 0.75 MIP\footnote{1 MIP corresponds to 80keV} energy after corrections for the particle's incident angle and for the pad thickness.
The definition of a hit is important as it changes the particle distribution itself in terms of both the total number of particles and the secondary/background ratio.
Since a single particle can produce hits in several pads, especially in high $\eta$ region in the Octagon, an $\eta$-$\phi$ cell map of size 0.05$\times$0.196 was introduced to allow merging of hits in that region.
Finally, an occupied cell in the $\eta$-$\phi$ map is considered as a hit for this analysis.

\vspace{-1cm}
\begin{figure}[h]
\begin{minipage}{7.5cm}
\mbox{\epsfig{file=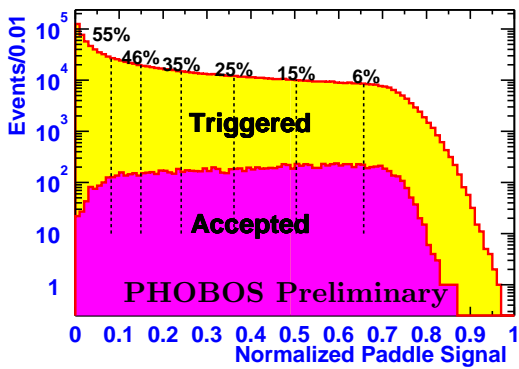,width=8cm}} 
\put(-60,10){\mbox{\small \bf PHOBOS Preliminary}}
\end{minipage}
\hspace{1cm}
\begin{minipage}[c]{7.5cm}
\caption{The distribution of the mean Paddle signal normalized by the maximum mean Paddle signal for triggered events (upper curve) and for data accepted in the final analysis (lower curve). The dashed lines delineate the centrality bins used in the analysis. The numbers on the lines in each bin represent the integrated fraction of triggered events.}
\label{fig:pm}
\end{minipage}
\end{figure}
\vspace{-1cm}

The analysis presented here is based on the data taken during the first RHIC physics run in 2000.
1.3 million events were accepted after checks of the detector performance and the vertex quality cut.
A narrow fiducial cut was applied on the $Z$ vertex position (-38cm$<z_{vtx}<$-30cm) in order to ensure that the Octagon covers $-2<\eta<2$ with a uniform and symmetric acceptance.
The final sample used in the analysis amounts to 13K events.
The selected data are divided into 7 bins of centrality using the paddle mean signal as shown in Fig.~\ref{fig:pm}.  

\section{Analysis method}
The formalism for the flow analysis presented here is described in detail in \cite{ref:p&v}.
First, the event plane is reconstructed by the anisotropic flow itself by measuring the final state flow vector, $\vec{Q}=(X_n,Y_n)=(\sum_i w_i \cos(n\phi_i),\sum_i w_i\sin(n\phi_i))$, where the sum runs over all particles, $i$, in an event and $w_i$ are the weights to optimize the event plane measurement. 
The event plane angle $\Psi_n$ is then simply given as $\tan^{-1}(X_n/Y_n)/n$.
Second, the observed flow signal is calculated as the coefficient of the Fourier expansion with respect to the event plane, $v_n^{obs}=<\cos(n(\phi-\Psi_n))>$. 
The actual flow signal $v_n$ measured with respect to the real reaction plane is then evaluated by taking into account the resolution in the event plane determination: $v_n=v_n^{obs}/<\cos(n(\Psi_n-\Psi_R))>$.
The event plane resolution can be directly calculated from the data using a sub-event technique since the correlation between two equal multiplicity sub-events yields a relation of $<\cos(n(\Psi_n^{a,b}-\Psi_R))>=\sqrt{ <cos(n(\Psi_n^a-\Psi_n^b))>}$, where $a$ and $b$ denote each sub-event.

\begin{figure}[h]
\begin{minipage}{7.5cm}
\mbox{\epsfig{file=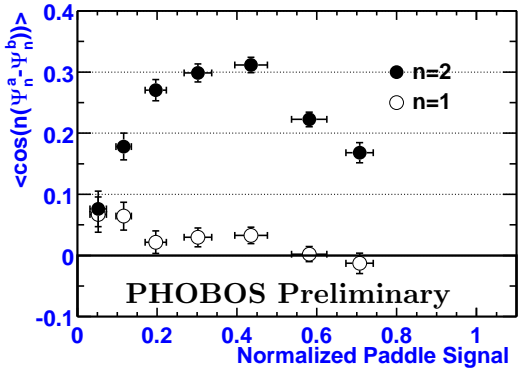,width=8cm}}
\put(-60,10){\mbox{\small \bf PHOBOS Preliminary}}
\end{minipage}
\hspace{1cm}
\begin{minipage}{7.5cm}
\caption{Correlations between two event plane angles determined from the two sub-events: one from the negative pseudo-rapidity sub-event and one from the positive pseudo-rapidity sub-event.}
\label{fig:corr}
\end{minipage}
\end{figure}

In this analysis, sub-events are defined by dividing particles into two pseudo-rapidity regions: one from -2.0 to -0.1 and the other from 0.1 to 2.0.
The pseudo-rapidity gap between the two sub-events is introduced to reduce the short-range correlations between the sub-events. 
Detector related biases such as cracks between sensors, variation in $\phi$ acceptance of pads, and asymmetric production of background particles, are removed by taking the pad weights, $w$, as the inverse of the average particle density at each pad integrated over all events.
After applying the weight correction, the $\Psi_{n}$ distributions become uniform.
Fig.~\ref{fig:corr} shows the correlations between the sub-event plane angles.
It can be seen that this analysis is quite sensitive to an elliptic flow $v_2$ measurement, while the statistics of the sample are insufficient for a meaningful determination of $v_1$. 

\vspace{-0.8cm}
\vspace{-1cm}

The second Fourier coefficients $v_2^{raw}(=<\cos(2(\phi^a - \Psi^b))>)$ are evaluated by the azimuthal angle distribution of hits in one sub-event $(\phi^a)$ with respect to the event plane from the other sub-event $(\Psi^b)$.
We use the notion of $v_2^{raw}$ because it is extracted from the $\eta$-$\phi$ cell map, not directly from the particle angular distribution, and therefore this is not necessarily equal to $v_2^{obs}$ as defined above.
Monte Carlo (MC) simulation studies were performed to understand how $v_2^{raw}$ is correlated to $v_2^{obs}$.
It was found that $v_2^{raw}$ decreases relative to $v_2^{obs}$ due to the saturation of the detector with hits.
The degree of the $v_2$ signal suppression, v$_{2}^{raw}$/v$_{2}^{obs}$, could be sensitive to the detector occupancy and the magnitude of the flow itself.
Extensive studies of simulated data show that the suppression has a negligible dependence on the size of flow for $v_2<10\%$, and can be parameterized by $(1-Occ)$, where $Occ$ is the fraction of occupied cells in the $\eta$-$\phi$ map.
Thus the observed flow value is deduced as $v_{2}^{obs} = v_2^{raw}/ (1-Occ)$.
The final elliptic flow strength, $v_{2}$, is derived by correcting $v_{2}^{obs}$ with the sub-event plane resolution as described previously.

\section{Results}

\vspace{-1cm}
\begin{figure}[h]
\begin{minipage}{7.5cm}
\mbox{\epsfig{file=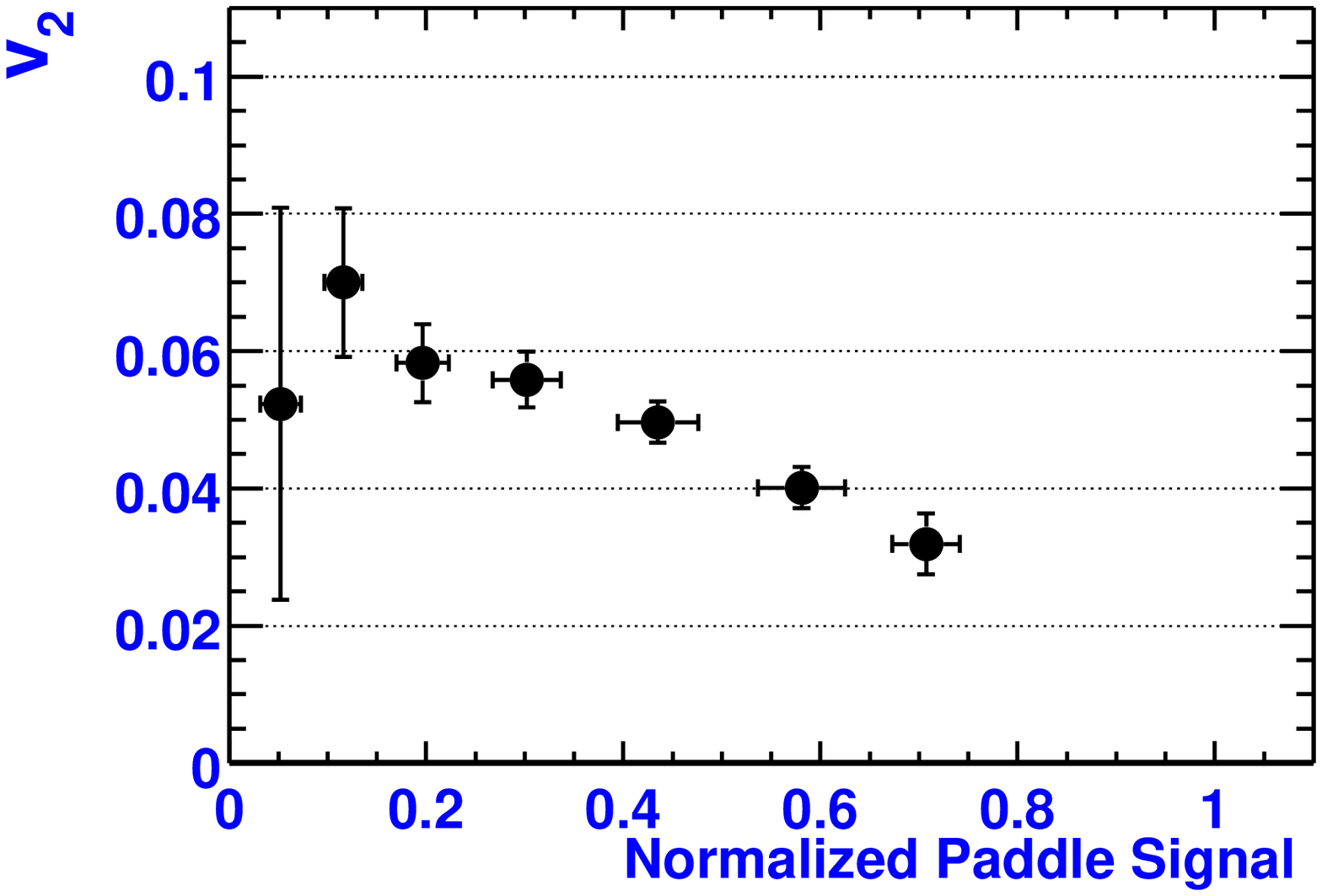,width=8cm}} 
\put(-60,40){\mbox{\small \bf PHOBOS Preliminary}}
\vspace{-1.2cm}
\caption{Elliptic flow, $v_2$, in the region -1.0$<\eta<$1.0 as a function of centrality. The error bars are statistical only.}
\label{fig:v2cen}
\end{minipage}
\hspace{1cm}
\begin{minipage}{7.5cm}
\mbox{\epsfig{file=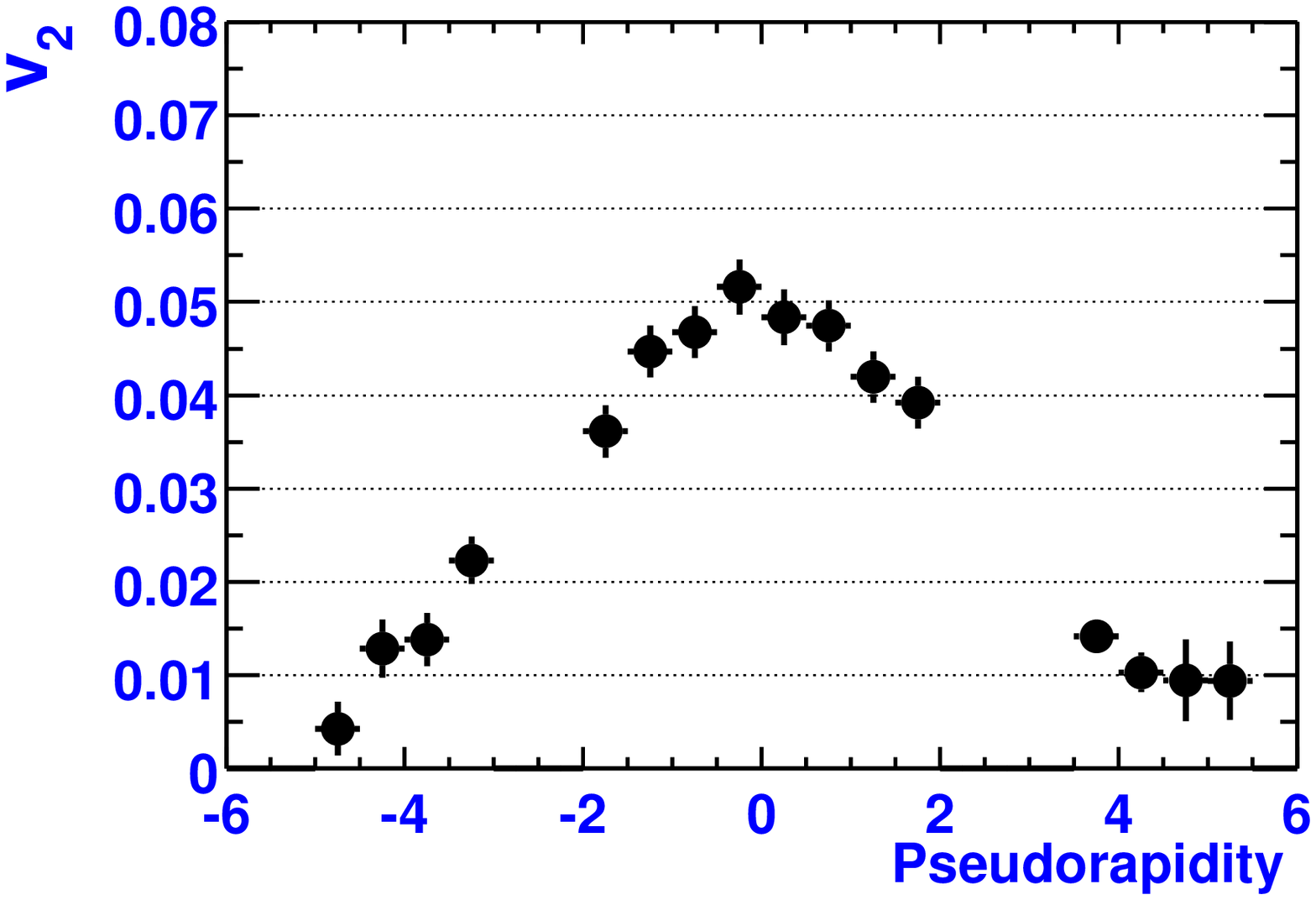,width=8cm}} 
\put(-60,40){\mbox{\small \bf PHOBOS Preliminary}}
\vspace{-1.2cm}
\caption{Elliptic flow, averaged over centrality, as a function of
$\eta$. The error bars are statistical only} 
\label{fig:v2eta}
\end{minipage}
\end{figure}
\vspace{-.5cm}

The centrality and pseudo-rapidity dependences of $v_2$ are shown in Fig.~\ref{fig:v2cen} and Fig.~\ref{fig:v2eta}, respectively.
A strong centrality dependence of $v_2$ ranging from 0.03 for the most central events up to 0.07 for peripheral events is observed.
The centrality dependence generally follows the results of a Hydrodynamic Model calculation. 
This is a possible indication that a high degree of thermalization is achieved in the early collision system by a large number of rescatterings.
The present experimental result is in excellent agreement with an earlier measurement from the STAR Collaboration~\cite{ref:snelling}.
The measurement of $v_2$ at high pseudo-rapidity, showing a similar shape to $dN/d\eta$, is a unique result from PHOBOS.

Various cross-checks were done using MC simulation to confirm the analysis method by studying the consistency between input flow and the resulting value.
An overall systematic error of 0.007 is estimated from studies of numerous sources of systematic uncertainty: $(1-Occ)$ parameterization, hit definition, beam orbit, relative alignment between Octagon, Rings, and Spectrometers.
Systematic errors related to the vertex algorithm, weighting matrix, sub-event technique, and magnetic field settings, were all found to be negligible.  

{
Acknowledgements: 
This work was partially supported by: (US) DoE grants DE-AC02-98CH10886,
DE-FG02-93ER40802, DE-FC02-94ER40818, DE-FG02-94ER40865,
DE-FG02-99ER41099 and W-31-109-ENG-38,
NSF grants 9603486, 9722606 and 0072204,
(Poland) KBN grant 2 P03B 04916,
(Taiwan) NSC contract NSC 89-2112-M-008-024.
}

\end{document}